\documentclass[aps,prl,preprint,superscriptaddress,floatfix,longbibliography]{revtex4-1}
\usepackage[bookmarks=true,colorlinks,citecolor=blue,urlcolor=blue]{hyperref}
\usepackage{ucs}
\usepackage[latin1]{inputenc}
\usepackage[english]{babel}
\usepackage[usenames, dvipsnames]{xcolor}
\usepackage{amsmath, 
		amsfonts,
           	booktabs,
            	microtype,
            	mathtools, 
            	graphicx,
            	xcolor,
            	nicefrac,
            	dsfont,
            	url, 
            	ulem,
            	placeins,
            	braket,
            	hyperref,
		appendix, 
		amssymb,
		soul}

\let\oldcite\cite
\renewcommand{\cite}[1]{\mbox{\oldcite{#1}}}
\DeclareRobustCommand{\mb}[1]{%
 \ifmmode\text{\setulcolor{red}\ul{$#1$}}\else\setulcolor{red}\ul{#1}\fi
}

\newcommand{\eq}[2]{\begin{eqnarray}\label{#1} #2 \end{eqnarray}}

\begin{document}

\title{Signatures of Self-Organised Criticality in an Ultracold Atomic Gas}
\author{S. Helmrich}
\affiliation{Physikalisches Institut, Universit\"at Heidelberg, Im Neuenheimer Feld 226, 69120 Heidelberg, Germany}
\author{A. Arias}\affiliation{Physikalisches Institut, Universit\"at Heidelberg, Im Neuenheimer Feld 226, 69120 Heidelberg, Germany}\affiliation{IPCMS (UMR 7504) and ISIS (UMR 7006), University of Strasbourg and CNRS, 67000 Strasbourg, France}
\author{G. Lochead}\affiliation{Physikalisches Institut, Universit\"at Heidelberg, Im Neuenheimer Feld 226, 69120 Heidelberg, Germany}\affiliation{IPCMS (UMR 7504) and ISIS (UMR 7006), University of Strasbourg and CNRS, 67000 Strasbourg, France}
\author{T. M. Wintermantel}\affiliation{Physikalisches Institut, Universit\"at Heidelberg, Im Neuenheimer Feld 226, 69120 Heidelberg, Germany}\affiliation{IPCMS (UMR 7504) and ISIS (UMR 7006), University of Strasbourg and CNRS, 67000 Strasbourg, France}
\author{M. Buchhold}\affiliation{Department of Physics and Institute for Quantum Information and Matter, California Institute of Technology, Pasadena, CA 91125, USA} 
\author{S. Diehl}\affiliation{Institut f\"ur Theoretische Physik, Universit\"at zu K\"oln, 50937 Cologne, Germany}
\author{S. Whitlock}\email[e-mail: ]{whitlock@unistra.fr}
\affiliation{Physikalisches Institut, Universit\"at Heidelberg, Im Neuenheimer Feld 226, 69120 Heidelberg, Germany}\affiliation{IPCMS (UMR 7504) and ISIS (UMR 7006), University of Strasbourg and CNRS, 67000 Strasbourg, France}
\date{\today}

\begin{minipage}[h]{\textwidth}
    \maketitle
\end{minipage}

\textbf{Self organisation provides an elegant explanation for how complex structures emerge and persist throughout nature. Surprisingly often, these structures exhibit remarkably similar scale-invariant properties. While this is sometimes captured by simple models that feature a critical point as an attractor for the dynamics, the connection to real-world systems is exceptionally hard to test quantitatively. Here we observe three key signatures of self-organised criticality in the dynamics of a driven-dissipative gas of ultracold atoms: (i) self-organisation to a stationary state that is largely independent of the initial conditions, (ii) scale-invariance of the final density characterised by a unique scaling function, and (iii) large fluctuations of the number of excited atoms (avalanches) obeying a characteristic power-law distribution. This establishes a well-controlled platform for investigating self-organisation phenomena and non-equilibrium criticality with unprecedented experimental access to the underlying microscopic details of the system.}

\begin{figure}[!t]
	\centering 
	\includegraphics[width = 0.62\columnwidth]{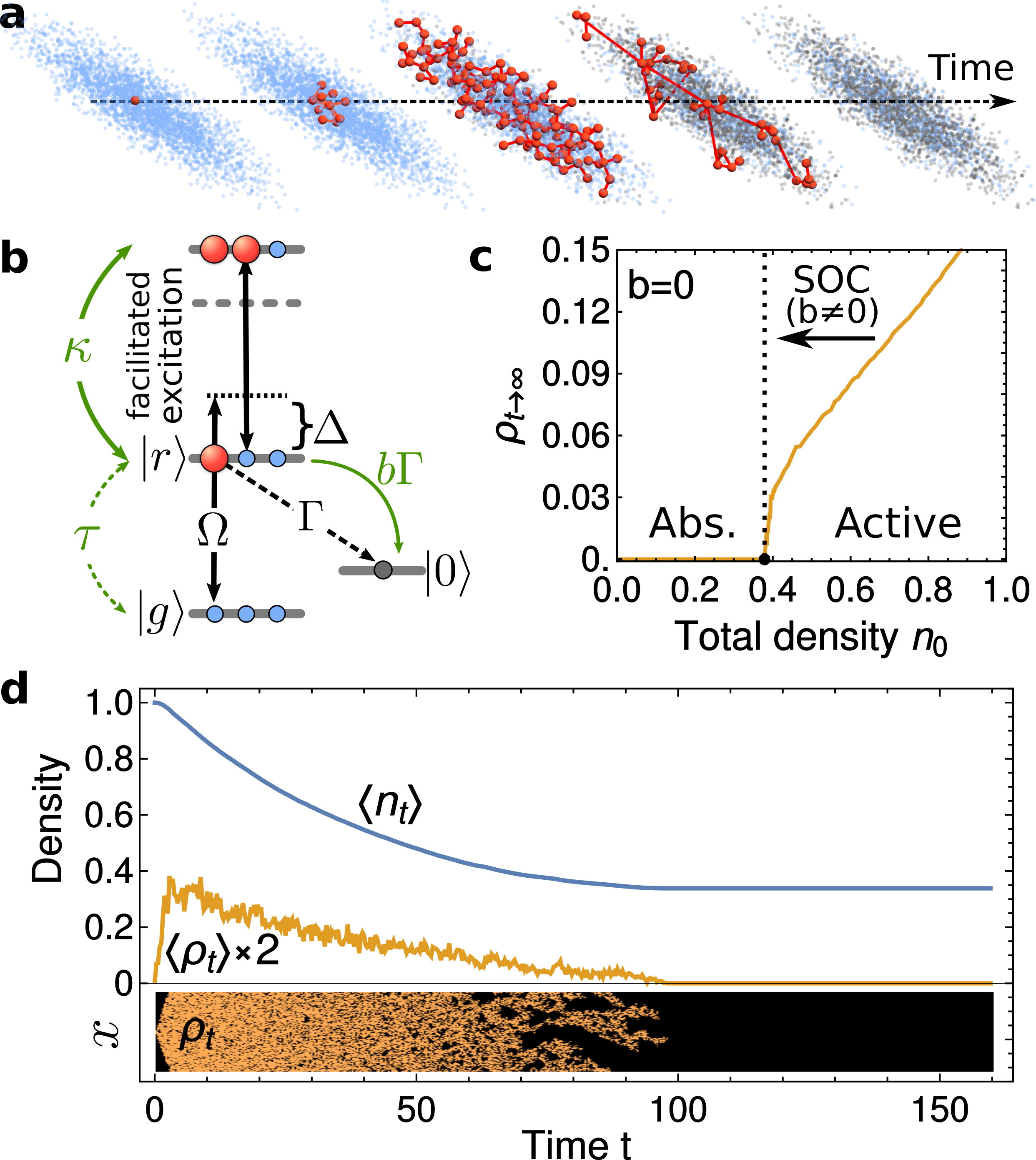}
	\caption{\textbf{Self-organised criticality in an ultracold atomic gas excited to Rydberg states by a laser field.} (a) Self-organisation process in a cigar shaped atom cloud showing atoms in the ground state $|g\rangle$ (blue dots) or excited to a Rydberg state $|r\rangle$ (large red spheres) via facilitated excitation processes leading to the build up of correlations (represented by red links). (b) The laser field couples the $|g\rangle \rightarrow |r\rangle$ transition with Rabi frequency $\Omega$ and detuning $\Delta$ while atoms in the $|r\rangle$ state either decay to removed states $|0\rangle$ (black circles) or facilitate further Rydberg excitations. These microscopic processes determine the couplings in the Langevin equation~(\ref{eq:langevin}) defined in the text (illustrated with green arrows and symbols). (c) Numerical solution of Eq.~(\ref{eq:langevin}) for the population conserving system $b=0$ (in one-dimension) with $D=1$ (discretisation distance = 1), $D_T=0$, $\Gamma=10$, $\kappa=10$ and $\tau=0$. As a function of the total density $n_0$, the stationary active density $\rho_{t\rightarrow\infty}$ exhibits an absorbing state phase transition (dotted vertical line) which acts as an attractor for the SOC dynamics (when $b\neq 0$). (d) Time evolution for $b=0.01$ showing the spatially-averaged active density $\langle\rho_t\rangle$ (orange) and total density $\langle n_t\rangle$ (blue) as the system approaches a stationary state close to the critical point of the absorbing state phase transition. The lower panel in (d) shows the full spatio-temporal evolution of the active density with transverse coordinate $x$.}
	\label{fig:1_temporal}
\end{figure}
Self-organised criticality (SOC) is a fascinating concept, first put forward by Bak, Tang and Wiesenfeld in 1987 as a way to explain the large abundance of scale-invariant systems found in nature~\cite{Bak1987}. It is thought to underlie a wide range of complex dynamical phenomena, ranging from activity in electrical circuits and neural networks~\cite{Hesse2014,Shew2015}, to the likelihood of avalanches and earthquakes \cite{Sornette1989} as well as how forest fires~\cite{Drossel1992,Malamud1998}, diseases~\cite{Rhodes1996} and even ideas spread~\cite{Gleeson2014}. However, despite the wide ranging fundamental and practical importance of the SOC phenomenon, much-needed clean experimental studies are hindered by numerous complexities concerning the relevant microscopic degrees of freedom~\cite{Field1995,Frette1996,dickman2000,Altshuler2004} and even the simplest toy models (beyond mean field approximations) present serious challenges to theory~\cite{Vespignani1997,Dickman1994,Munoz1996,Dornic2005,henkel2008non,Bonachela2009,Markovic2014}.

SOC can be understood as an organising principle governing a class of dissipative interacting systems that display three key signatures: (i) Self-organisation to a stationary state (bringing observables to values that are independent of initial conditions); (ii) Scale invariance of spatio-temporal correlation functions, including bulk observables; (iii) Critical response to small perturbations, usually encountered in the form of avalanches that have a broad range of sizes and durations described by power-law distributions. This is unlike at an equilibrium phase transition, where scale invariance and a critical response only ensues for a fine-tuned parameter set. The common root of these emergent SOC properties is that the respective gap (i.e. the distance in parameter space from the critical state) is replaced by a ``dynamical gap'' which self-tunes to zero by an intrinsic feedback mechanism. This, and the signatures (i)-(iii) set SOC apart from other occurrences of non-equilibrium scaling behaviour, such as hydrodynamic long-time tails~\cite{forster1990}, the Kosterlitz-Thouless critical phase in two-dimensional quantum fluids \cite{BKT,Halperin1978} and the transient dynamics of turbulent cascades in isolated systems~\cite{Frisch1995,Berges2008}, which have also been studied with ultracold atoms, see Refs.~\cite{Hadzibabic,Murthy2015} (Kosterlitz-Thouless phase) and~\cite{Tsatsos2016,Navon2018,Prufer2018,Erne2018} (turbulence), as well as related experiments on superradiance~\cite{Inouye1999, Clark2017} and scaling in unitary Bose gases~\cite{Ho2004}.

In this work, we demonstrate the phenomenology of SOC [signatures (i) to (iii)] in the dynamics of a microscopically well-controlled physical system consisting of a three-dimensional trapped gas of ultracold potassium atoms driven to highly excited Rydberg states by a laser field (Fig.~\ref{fig:1_temporal}a). As we will show, the crucial new ingredient leading to SOC is the slow irreversible decay of the excited population to auxiliary inactive states, which has thus far been largely disregarded in the investigation of Rydberg many-body dynamics. This enables the observation of a phase transition from a self-organising active phase to an absorbing phase, scale-invariance of the self-organised density and large fluctuations of the active density that take the form of power-law distributed avalanches. Beyond these experimental results, we derive a Langevin equation from the underlying microscopic many-body quantum master equation governing driven-dissipative Rydberg dynamics, which coincides with one of the emblematic classes of SOC models~\cite{Bonachela2009}. This provides the crucial link from the microscopic atomic physics to the observed macroscopic SOC phenomenology and establishes ultracold Rydberg atomic gases as a widely tunable and theoretically accessible platform for the study of self-organisation phenomena and universality in non-equilibrium dynamics.

\textit{Physical system} -- Each of the approximately $10^5$ atoms held in the optical trap can be represented by a three state system comprising: the ground state $|g\rangle=|4s_{1/2},F=1\rangle$, an excited Rydberg state $|r\rangle$, and auxiliary removed states, which we refer to collectively by $|0\rangle$ (Fig.~\ref{fig:1_temporal}b). The laser field continuously drives the $|g\rangle\rightarrow |r\rangle$ transition with a fixed detuning $\Delta$ from resonance and with an amplitude parameterised by the Rabi frequency $\Omega$. In our experiments $\Delta\gg\Omega$ such that spontaneous single-atom excitation processes are strongly suppressed. Once excited however, atoms can facilitate further excitations (e.g. when the laser detuning is compensated by the interaction energy of Rydberg pair states) leading to the formation of extended excitation clusters~\cite{Lesanovsky2013,Carr2013,Schempp2014,Malossi2014,Urvoy2015,Valado2016,Goldschmidt2016,Simonelli2016,Letscher2017,Gutierrez2017}. Alternatively, they can be spontaneously lost from the system, predominantly by decaying to another hyperfine ground state that is not coupled by the lasers or to other states that are not optically trapped. 

One major advantage of this system is that it permits an effectively complete microscopic description in terms of a quantum master equation for the many-body density matrix $\hat{\rho}$
\begin{equation}\label{eq:me}
	\partial_t \hat{\rho}=\frac{i}{\hbar}[\hat{\rho},\hat{H}]+\sum_l\mathcal{L}_l(\hat{\rho})
\end{equation}
with atom-light interaction Hamiltonian $\hat{H}$ and Lindblad superoperator $\mathcal{L}_l(\hat{\rho})$ given by 
\begin{eqnarray}
	\hat{H}&=&\sum_l\Big[\Big(\sum_{l'}\frac{1}{2}\frac{C_6}{|{\bf{r}}_{ll'}|^6}\hat{\sigma}^{rr}_{l'}-\Delta\Big)\hat{\sigma}^{rr}_l+\frac{\Omega}{2}\big(\hat{\sigma}^{gr}_l+\hat{\sigma}^{rg}_l\big)\Big],\nonumber \\
	\mathcal{L}_l(\hat{\rho})&=&\Gamma\hat{\sigma}^{0r}_l\hat{\rho}\hat{\sigma}^{r0}_l+\gamma_\mathrm{{de}}\hat{\sigma}^{rr}_l\hat{\rho}\hat{\sigma}^{rr}_l-\frac{\gamma_\mathrm{{de}}+\Gamma}{2}\Big( \hat{\sigma}^{rr}_l \hat{\rho}\ + \hat{\rho} \hat{\sigma}^{rr}_l \Big), \nonumber 
\end{eqnarray}
where $\hat{\sigma}^{\alpha\beta}_l\equiv |\alpha\rangle\langle\beta |_l$ and $l,l'$ are indices for each atom. Interactions between Rydberg states are parameterised by the van der Waals coefficients $C_6/2\pi\approx 0.52\,\mathrm{GHz\,\mu m^6}$ for the $|r\rangle=|39p_{3/2}\rangle$ state and $C_6/2\pi\approx 238\,\mathrm{GHz\,\mu m^6}$ for the $|r\rangle=|66p_{3/2}\rangle$ state. Dissipation is described by $\mathcal{L}_l(\hat{\rho})$, which includes spontaneous loss (with total rate $\Gamma$) and irreversible dephasing (rate $\gamma_\mathrm{de}$) attributed primarily to residual laser phase noise and Doppler broadening.

To connect the microscopic dynamics of Rydberg atoms [Eq.~(\ref{eq:me})] to the SOC phenomenology, we apply a systematic coarse graining procedure for the collective dynamics (derived in the Supplementary Information). In brief, we average over the characteristic length scale corresponding to the facilitation process and project onto the density degree of freedom by adiabatically eliminating the rapidly decaying atomic coherences~\cite{Ates2007,Marcuzzi2014}. We also approximate the atomic medium as a quasi-homogeneous gas with a smoothly varying density, which is justified by the fact that the atoms are free to move on the timescale of the SOC dynamics. The final result is a Langevin equation for the density of atoms in the $|r\rangle$ state $\rho_t = \rho(t, \mathbf{r})$ (which we call the active component) and the total remaining density $n_{t}=n(t,{\bf r})$, which is the sum of the populations in the $|g\rangle$ and $|r\rangle$ states (excluding removed states):
\begin{eqnarray}
	\partial_t \rho_t &=&\  \big(D \nabla^2-\Gamma+\kappa n_t\big) \rho_t - 2 \kappa {\rho_t}^2 + \tau(n_t-2\rho_t)+ \xi_t,\nonumber \\
	n_t &=&\ n_0-b\Gamma\int_0^t \mathrm{d}t' \rho_{t'}+D_T\int_0^t \mathrm{d}t'\nabla^2 n_{t'}. 
	\label{eq:langevin}
\end{eqnarray}
In this equation $D$ and $D_T$ are the diffusion constants for the active and total densities respectively and $\kappa$ is the facilitation rate which together govern the rate of excitation spreading, $\tau$ is the spontaneous excitation rate, $n_0$ is the initial density, and $b$ is a dimensionless parameter governing how fast decay depletes the total population. The stochastic part of the evolution is governed by the uncorrelated multiplicative noise term $\xi_t=\xi(t, \mathbf{r})$ with variance $\mathrm{var}(\xi_t)=\Gamma \rho_t$. 

Eq.~(\ref{eq:langevin}) closely coincides with the paradigmatic Drossel-Schwabl forest fire model~\cite{Drossel1992,Bonachela2009}, except for the absence of a slow regrowth term for the total density, which would normally bring the system from an inactive (subcritical) state to the critical state. This regrowth is typically the slowest scale in the model and must asymptotically vanish in order to realise SOC. However in the absence of regrowth the system still exhibits a non-equilibrium phase transition~\cite{Bonachela2009}. This highlights an important advantage of the ultracold atom system, since one can readily prepare the system initially in the active phase and observe the approach to the SOC state. Thus Eq.~(\ref{eq:langevin}) captures all essential features of SOC, which could be observed in experiments. To illustrate this we present in Figs.~\ref{fig:1_temporal}(c,d) numerical simulations, for simplicity focusing on a small one-dimensional system. In the case: $b = \tau = 0$, the system features a non-equilibrium phase transition~\cite{Gutierrez2017} from an absorbing phase, where any excited component quickly dies out (characterised by $\rho_{t\rightarrow\infty}\rightarrow 0$ for $\kappa n_0 \ll \Gamma$), to an active phase in which excitations spread throughout the system from arbitrarily small seed excitations (with $\rho_{t\rightarrow\infty}>0$ for $\kappa n_0 \gg \Gamma$). For $b,\tau \neq 0$ on the other hand, spontaneous single-atom excitations trigger the relatively fast facilitated excitation dynamics, while on longer timescales particle loss introduces a coupling between $\rho_t$ and $n_t$. Specifically, the first integral in Eq.~(\ref{eq:langevin}) acts as a feedback mechanism, causing $n_t$ to continuously decrease while in the active phase. When this loss is much slower than the internal dynamics but much faster than the spontaneous excitation rate (achieved for the separation of timescales $\kappa n_0 \sim \Gamma \gg b \Gamma \gg \tau$), the system slowly approaches the underlying critical point of the absorbing state phase transition and develops scale invariant properties, witnessed for example by growing spatio-temporal correlations in the active density (fractal-like structures seen around t=80 in the lower panel of Fig.~\ref{fig:1_temporal}d). This behaviour can be understood in terms of the evolution of the dynamical gap $\kappa n_t-\Gamma$, which is initially positive and continuously decreases due to population loss, until it asymptotically reaches zero at the critical point, where the dynamics effectively stop.

Theoretically solving the full dynamical many-body problem described by Eq.~(\ref{eq:langevin}) beyond the mean-field level is formidably difficult, particularly for large system sizes and in more than one spatial dimension, due to the presence of multiplicative noise and the importance of strong spatio-temporal correlations~\cite{Dickman1994,Dornic2005}. As a result, many properties of this class of systems are still actively debated, such as the crucial question of whether the system self-organises towards a truly critical state, and whether it fulfils universal scaling relations conjectured for SOC~\cite{Munoz1996,Bonachela2009}.
In particular, the non-equilibrium critical exponents for the model described by Eq.~(\ref{eq:langevin}) have not been reliably determined beyond the mean field level. Limiting cases have been fully addressed, instead. For example, in the case that $b=\tau=0$ (number conserving and Markovian limit) the critical behaviour is governed by a critical point in the directed percolation universality class~\cite{henkel2008non,Gutierrez2017}. How this universality changes in the non-commuting limit of $b$ being small but non-vanishing is not conclusively understood~\cite{Munoz1996,Bonachela2009}, but we expect it to be strongly modified since, in a renormalisation group picture, the fully attractive SOC fixed point does not feature a relevant direction as is the case for directed percolation. In the non-interacting case $\kappa=0$, the model features a Gaussian fixed point (with respect to $\kappa$) in the dynamical percolation universality class~\cite{henkel2008non}. However, this is presumably not smoothly connected to the Wilson-Fisher fixed point of the interacting non-Markovian problem. Even with state of the art numerical methods a precise determination of the critical properties represents a substantial challenge~\cite{Dickman1994,Dornic2005}. In what follows, we experimentally implement this elusive model and and provide a first experimental characterisation of some of its scale invariant properties as the system approaches the non-equilibrium critical point.

\begin{figure}[!t]
	\centering 
	\includegraphics[width = 0.76\columnwidth]{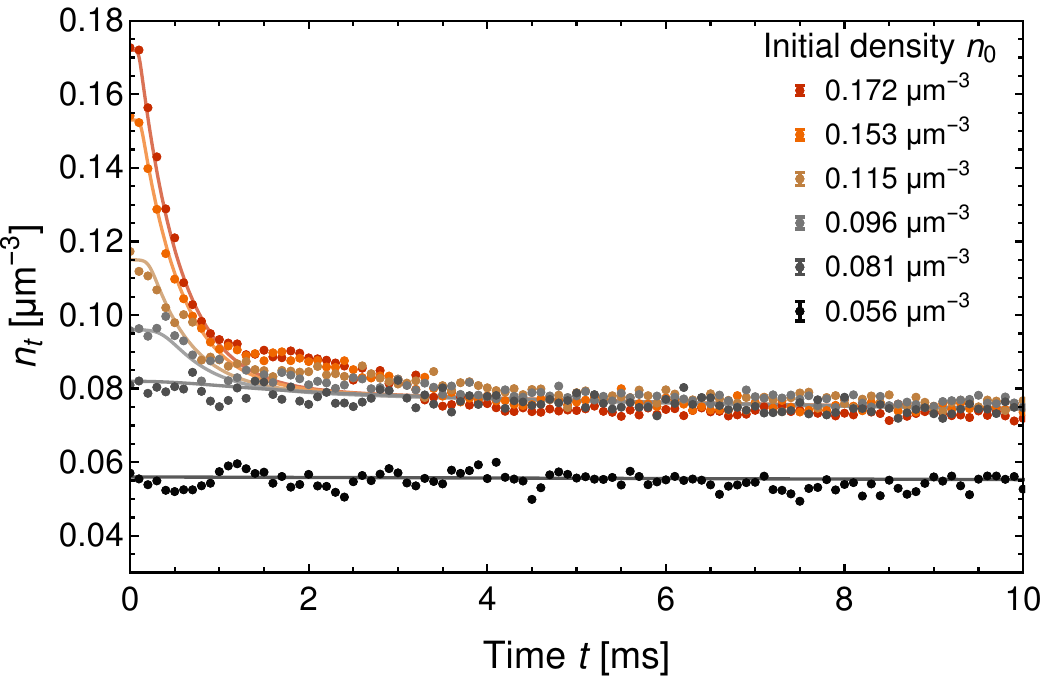}
	\caption{\textbf{Self-organisation: Above a threshold value, the remaining total atom density $n_t$ is attracted to the same stationary state density independent of the initial conditions.} The Rydberg state used is $39p_{3/2}$ and the parameters of the driving laser field are $\Delta/2\pi=30\,$MHz, $\Omega/2\pi=190\,$kHz. For high initial densities $n_0\gtrsim 0.08\,\mu$m$^{-3}$ the time dependence consists of a short initial plateau followed by fast exponential decay to a unique stationary state with a fixed density $n_f=0.075\,\mathrm{\mu m^{-3}}$. For initial densities below $n_f$ the dynamics is effectively stationary (black points). The solid lines correspond to mean field solutions to the effective Langevin equation with parameters given in the text. Standard errors for each dataset are indicated by the representative errorbars shown in the legend at top right.}
	\label{fig:2_verification}
\end{figure}

\textit{Self-organisation mechanism and model verification} -- We start our experiments by investigating the full time evolution of the total remaining density for different initial states. For this we prepare a gas of atoms in the atomic ground state ($\rho_0=0$) with different initial peak atomic densities $n_0$ between $0.056(5)\,\mathrm{\mu m^{-3}}$ and $0.172(2)\,\mathrm{\mu m^{-3}}$ where the numbers in parentheses refer to the standard error of the mean taken over several measurements. The Rydberg excitation laser is then suddenly switched on with constant $\Omega/2\pi=190\,$kHz and $\Delta/2\pi=30\,$MHz from the $39p_{3/2}$ state. After an adjustable time $t$ we turn off the excitation laser and then take an absorption image of the remaining atoms to determine $n_t$. Fig.~\ref{fig:2_verification} shows that the time evolution of $n_t$ is strikingly nonlinear, exhibiting two distinct types of behaviour depending on $n_0$. For high $n_0$ it starts with a short initial plateau, followed by a rapid exponential decay, reflecting the initial growth of the excitation density and a high degree of activity. This decay finally cuts off at a fixed density $n_f=0.075\,\mathrm{\mu m^{-3}}$ that is constant over a wide range of initial densities to a high accuracy (standard deviation of $0.003\,\mathrm{\mu m^{-3}}$), indicating a stable attractor for the many-body dynamics. In contrast, for $n_0<n_f$ the dynamics appears mostly frozen, characteristic of an absorbing phase. These two types of behaviour and the sudden transition in between signal the underyling absorbing state phase transition that depends upon the initial density and driving strength. On much longer timescales we observe a slower overall decay that we attribute to residual single atom excitations (and subsequent loss) with a characteristic rate $\tau/2\pi = 1.12(2)\,\mathrm{Hz}$. Because of this slow loss, the self-organised state is not sustained indefinitely, however the very large separation of timescales in our experiment makes it possible to robustly observe the emergent SOC features in the quasi-stationary regime (hereafter referred to as the stationary state).

We now verify that the Langevin equation provides a good theoretical description for the experimental observations. From this we confirm the required non-Markovian coupling between the active density and the total remaining density and the key hierarchy of scales $\kappa n_0 \sim \Gamma \gg b \Gamma \gg \tau$. For this it is sufficient to compare our data with a homogeneous mean field approximation to the Langevin equation ($D=0$ and $\xi_t=0$). We find that the mean field solutions, shown as solid lines in Fig.~\ref{fig:2_verification}, describe the data well, except for the minor deviation in the approach to the stationary state seen around $t\approx 2\,$ms. By simultaneously fitting all of the data shown with a single set of parameters we find $\Gamma/2\pi=11.7(9)\,\mathrm{kHz}$, $\kappa/2\pi=144(10)\,\mathrm{kHz\,\mu m^3}$ and $b=0.059(5)$, with statistical errors estimated using bootstrap resampling. Thus the required separation of scales is satisfied by an order of magnitude or more, placing our experiments firmly in the interesting regime for SOC. Furthermore, the experimental observations and theoretical confirmation establish the presence of the anticipated absorbing state phase transition and self-organisation to a stationary state which is independent of initial conditions [SOC signature (i)].

\begin{figure}[!t]
	\centering 
	\hskip -5pt\includegraphics[width = 0.76\columnwidth]{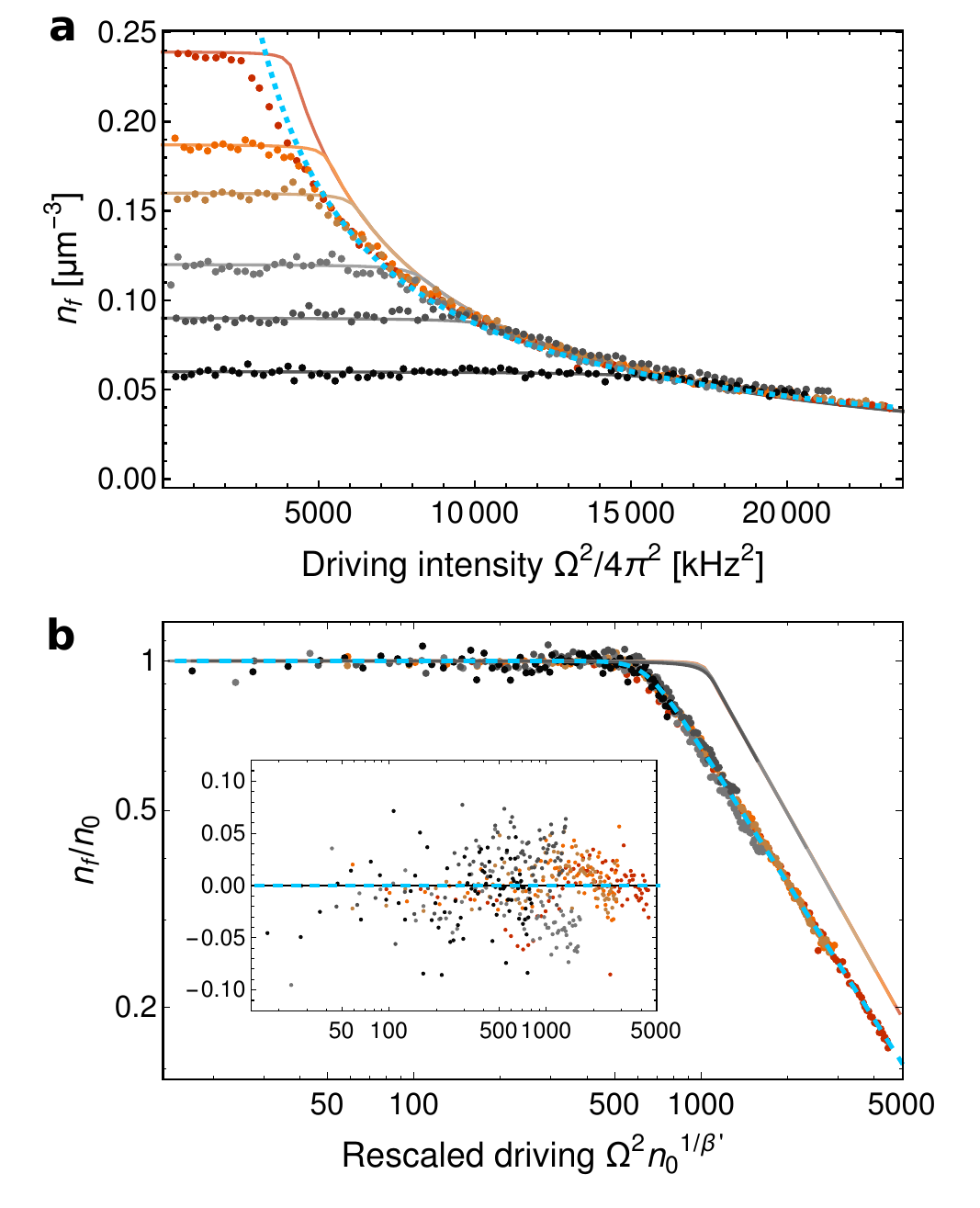}
	\caption{\textbf{Scale invariance of the self-organised stationary state as a function of the driving intensity $\Omega^2$.} (a) Stationary state density $n_f$ measured at $t = 10\,\mathrm{ms}$ as a function of $\Omega^2$ and for different initial densities $n_0$ using the same parameters as for Fig.~\ref{fig:2_verification} except $\Delta/2\pi=18\,$MHz. For large $\Omega^2$ and $n_0$ all points collapse onto one single powerlaw curve $n_f \propto \Omega^{-2\beta}$ (dotted blue line).  (b) The same data with rescaled axes to achieve full data collapse, revealing a unique scaling function (with fit shown by the dashed blue line) for the stationary density $n_f$. The inset shows the normalised residuals between the rescaled data and the fitted scaling function. Each datapoint is the mean of approximately 6 measurements. The solid lines in (a) and (b) correspond to mean field solutions of the effective Langevin equation.} 
	\label{fig:2_perceo}
\end{figure}

\textit{Scale-invariance of the stationary density} -- 
We now turn our attention to experimental manifestations of the observed phase transition on the stationary state. For this, the parameter $\kappa$ is of special importance, since it determines the critical point of the absorbing state phase transition and its corresponding scaling behaviour. In our system $\kappa$ is determined microscopically by the Rydberg excitation rate proportional to the driving intensity $\Omega^2$ which can be tuned over a wide range. By comparing experimental timetraces for different detunings we additionally infer the relation $\kappa \propto \Omega^2/\Delta^{2.06(1)}$, which we use to compare the following experimental results with theory without further free parameters. In Fig.~\ref{fig:2_perceo} we examine the dependence of the stationary density $n_f$ (reached after $t=10\,$ms of evolution) on the driving intensity. For different initial densities $n_0$, the stationary state exhibits a clear density dependent critical intensity $\Omega_c^2$ separating the absorbing phase (with $n_f \approx n_0$) from the active self-organising phase (with $n_f<n_0$). For the latter, the data falls onto a single curve resembling a power-law that is independent of the initial density (dotted blue line in Fig.~\ref{fig:2_perceo}a). While mean field theory (solid lines) reproduces the qualitative features of this data, the real, three dimensional system exhibits important quantitative differences that go beyond the mean field prediction, including a shift in the threshold intensity seen for higher initial densities and a markedly different powerlaw exponent. 

To further quantify the scale-invariant properties, we apply the scaling ansatz\break $n_f=\nobreak n_0 F(\Omega^2 n_0^{1/\beta'})$~\cite{henkel2008non}. By plotting $n_f/n_0$ as a function of $n_0^{1/\beta'}\Omega^2$, all of the data collapse onto a single universal curve (Fig.~\ref{fig:2_perceo}b), with the best results obtained with a rescaling exponent $\beta'=0.869(6)$. We find that the scaling function $F(x)$ can be very well modelled by the heuristic function $F(x)=x_c^\beta(x^{v\beta}+x_c^{v\beta})^{-1/v}$ (dashed blue curve in Fig.~\ref{fig:2_perceo}b), where $x_c$ and $v$ are free parameters describing the position and sharpness of the transition region between absorbing and active phases. For $x\gg x_c$ the scaling ansatz is a powerlaw $n_f\sim\Omega^{-2\beta}n_0^{1-\beta/\beta'}$, therefore we identify $\beta$ as the scaling exponent which characterises the stationary density and $1-\beta/\beta'$ quantifies how (in)sensitive it is to the initial density. Fitting the rescaled data on a log-log scale we obtain $\beta=0.910(4), v=10.6(8)$ and $x_c=641(3)\,\mathrm{kHz^2 \mu m^{-3/\beta'}}$ where the small statistical errors of the fitted parameters are indicated in parentheses. Using these parameters, the scaling function describes the stationary density extremely well. This is confirmed by the small and evenly scattered normalised residuals between the rescaled data and the fitted scaling function, spanning both the absorbing and active phases (Fig.~\ref{fig:2_perceo}b-inset). The clean power-law dependence additionally rules out substantial modifications due to the finite system size or inhomogeneous trapping geometry (see also Supplementary Information). Additional data taken for different densities and detunings of the driving field and slightly different experimental conditions exhibits a very similar scaling form and confirms this measured scaling exponent within an accuracy of a few percent (Supplementary Information). In contrast, the mean-field scaling solution predicts $\beta'_{MF}=\beta_\mathrm{MF}=1$, which is clearly incompatible with our data. While it is still debated to what extent SOC systems exhibit clean universal behaviour~\cite{dickman2000,Bonachela2009,Watkins2016}, it is striking that a single function describes the stationary state over the entire accessible parameter regime and that this function acquires a scale-invariant form characterised by a non-trivial scaling exponent [confirming SOC signature (ii)].
 
\textit{Power-law distributed excitation avalanches} -- Having established the scale-invariant nature of the self-organised stationary state through the remaining density, we now show that the SOC state is also evident in the statistical fluctuations of the active component. For this we use a different detection method based on field ionisation of the Rydberg excited atoms. This method is sensitive down to a few individual excitations, but in our experiment could only be applied for Rydberg states with principal quantum numbers $n>60$ due to the electric field dependence of the field ionisation threshold. Therefore, for the following measurements we use the $66p_{3/2}$ state, but otherwise comparable experimental conditions to the previous results. The measurement is also destructive, therefore each measurement point corresponds to a new experimental realisation. 

\begin{figure}[t!]
	\includegraphics[width = 0.90\columnwidth]{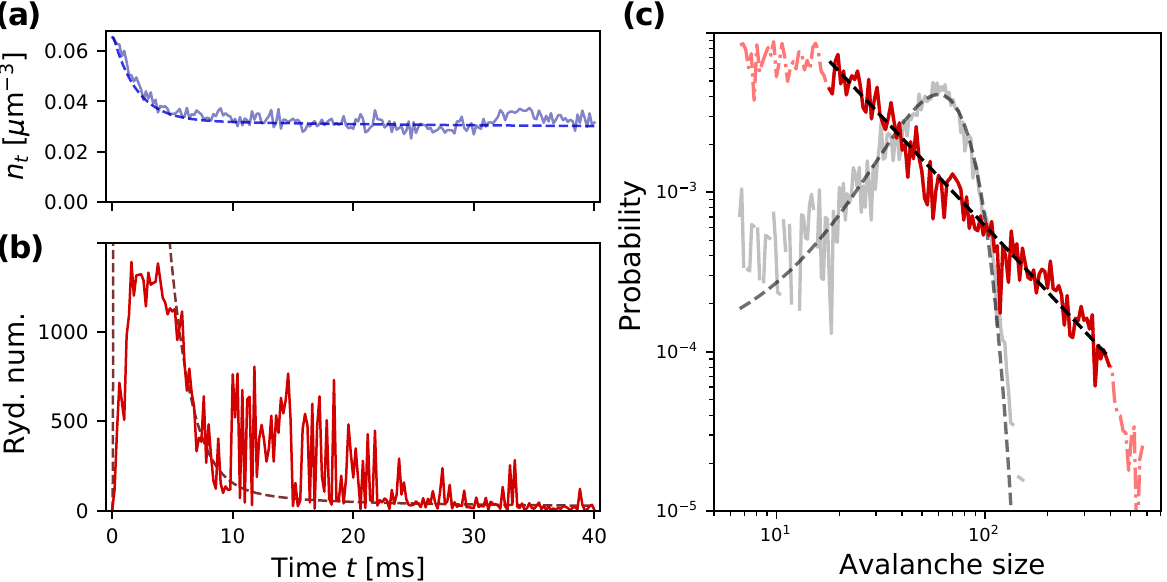}
	\caption{\textbf{Observation of power-law distributed excitation avalanches.} (a) Evolution of the remaining density for the $66p_{3/2}$ state. (b) Simultaneously measured Rydberg atom number (active component) integrated over the whole cloud showing large fluctuations of the active density for $t\gtrsim 10\,$ms. Each point corresponds to a new realisation of the experiment. The dashed lines in (a) and (b) are mean-field predictions, where the effective cloud volume in (b) is adjusted for optimal agreement. (c) Probability distribution for the instantaneous number of Rydberg excitations (avalanche size) for 3630 experimental runs. To determine the power-law exponent we truncate the data in a finite window indicated by the solid red line and apply maximum-likelihood estimation. The powerlaw exponent $\alpha=1.37(2)$ is depicted by the straight dashed line. The gray data corresponds to a control measurement for resonant excitation with a short duration laser pulse which does not exhibit a power-law distribution.}
	\label{fig:powerlaw}
\end{figure}

Figure~\ref{fig:powerlaw}a shows a time trace of the temporal evolution of the remaining density and the instantaneous number of excitations (active component). The remaining density follows the same characteristic non-exponential time dependence as seen in Fig.~\ref{fig:2_verification} except for overall slower dynamics that can be explained by the longer lifetime and larger $C_6$ coefficient for the $66p_{3/2}$ state, which lowers $b$ in the effective description (Supplementary Information). Fig.~\ref{fig:powerlaw}b shows that the active component undergoes a rapid growth at early times which saturates the detector around $2-5$\,ms and then reduces again as a consequence of the associated fast atom loss. After $10\,$ms the remaining density has almost reached is stationary value, yet we observe very large fluctuations of the excitation number ranging from almost zero to clusters of up to $\approx 800$ excitations. We interpret this as the strong response of the system to individual excitation events that trigger the formation of large excitation clusters that spread across the system with a broad distribution of sizes and durations (avalanches), expected as the dynamical gap vanishes close to the critical point [SOC signature (iii)]. In Supplementary Fig.~\ref{fig:3_sus} we present additional evidence of this strong response in the bulk observables following a parameter quench. The solid lines in Fig.~\ref{fig:powerlaw}a,b show the mean-field solution to the effective Langevin equation, which describes the remaining atom number quite well, but as expected completely fails to capture the large fluctuations. Additionally we observe avalanches over a wide time window (up to 40\,ms) even though the remaining density appears mostly stationary. This shows that the system remains close to the self-organised critical state for an extended time period, despite the absence of an obvious particle reloading mechanism which would be required to keep the system at the critical point indefinitely.

To investigate the distribution of avalanche sizes $s$, we chose a fixed time of $25\,$ms and repeated the experiment $3630$ times. At this time the observed excitation spikes are relatively sparse (enabling their interpretation of individual avalanche events), yet they occur frequently enough to obtain sufficient statistics. Figure~\ref{fig:powerlaw}c shows the corresponding empirical probability distribution function obtained by binning the data using logarithmically spaced intervals and plotted on a double logarithmic scale. The empirical probability distribution function is well described by a power-law spanning $1.5$ orders of magnitude with an upper cutoff determined by the finite system size or detector saturation (both effects are expected to play a role around $s \gtrsim 500$). The plateau around $s\lesssim 20$ is attributed to the noise floor of the detector. To confirm that the observed power-law distribution is indeed a feature of the self-organising dynamics, we also show in Fig.~\ref{fig:powerlaw}c a comparable distribution obtained by a resonant excitation pulse of $1$~$\mu$s duration which yields a stretched Poissonian distribution as expected for mostly uncorrelated excitations. %
To estimate the power-law exponent we truncate the empirical data in the window $20\leq s\leq 400$ (corresponding to $2450$ measurements), and apply maximum likelihood estimation~\cite{Bauke2007}, yielding a power-law exponent of $\alpha = -1.37(2)$, where the statistical uncertainty was estimated using bootstrap resampling. We also confirm via Kolmogorov-Smirnov testing that the power-law hypothesis is favoured over other common distributions such as exponential, log-normal, and gamma distributions. The power law exponent falls in a similar range to observations made in other conjectured SOC-like systems, such as forest-fires~\cite{SONG200161}, neuronal networks~\cite{Kaus2011}, earthquakes and solar flares~\cite{Arcangelis2006}. It is important to note however that non-universal corrections, e.g., due to the non-vanishing dissipation and driving rates or imperfect separation of scales, could still have an effect on the apparent critical properties~\cite{Bonachela2009}. Another advantage of the ultracold atom platform therefore is the prospect to control these experimental conditions (e.g. larger detunings correspond to lower seed excitation rates) and to determine the critical exponents for different dimensionalities in a single experimental system, permitting more stringent tests of universal scaling predictions.

\textit{Conclusion} -- In this work we identified self-organised criticality as the mechanism governing the dynamics of a driven-dissipative, open many-body system. This is observed through the strikingly nonlinear evolution of the system to a stationary state which is independent of the initial conditions and exhibits scale-invariance over a wide range of initial densities, driving strengths and detunings as well as excitation avalanches with a broad distribution of sizes characteristic of a power-law distribution. This is further supported by the theoretical model that links the microscopic physical processes to paradigmatic SOC models that will permit quantitative theory-experiment comparisons beyond the mean-field level. The demonstrated versatility of ultracold Rydberg gases combined with the ability to understand and experimentally control the microscopic physics in this system almost completely makes it a unique platform for studying non-equilibrium collective behaviour. Future experiments could implement a mechanism to slowly add particles to the system [i.e. an additional regrowth term in Eq.~(\ref{eq:langevin})] to sustain the SOC state on even longer timescales. It should also be possible to investigate other observables, including spatio-temporal correlations in the active and remaining densities. This would make it possible to determine multiple critical exponents and scaling relations which would help answer long standing questions about the universal or non-universal aspects of SOC and its relation to other non-equilibrium universality classes. Additionally, experiments could explore the interface between driven-dissipative and isolated quantum systems governed by competing classical and quantum dynamical rules~\cite{Marcuzzi2016,Buchhold2017,PerezEspigares2017}, ultimately leading to a more complete and quantitative understanding of non-equilibrium universality.

\acknowledgements{We acknowledge Thomas Ebbesen, Guido Pupillo and Matthias Weidem\"uller for discussions. This work is supported by the Deutsche Forschungsgemeinschaft under WH141/1-1 and is part of and supported by the DFG Collaborative Research Centre ''SFB 1225 (ISOQUANT)'',  the Heidelberg Center for Quantum Dynamics, the European Union H2020 FET Proactive project RySQ (grant N. 640378) and the `Investissements d'Avenir' programme through the Excellence Initiative of the University of Strasbourg (IdEx). M. B. acknowledges support from the Alexander von Humboldt foundation. S.D. acknowledges support by the German Research Foundation (DFG) through the Institutional Strategy of the University of Cologne within the German Excellence Initiative (ZUK 81) and the European Research Council via ERC Grant Agreement n. 647434 (DOQS). S.W was partially supported by the University of Strasbourg Institute for Advanced Study (USIAS), S.H. acknowledges support by the Carl-Zeiss foundation, A.A. and S.H. acknowledge support by the Heidelberg Graduate School for Fundamental Physics.} 

{\center\textbf{METHODS}}\\

\noindent\textbf{A. Sample preparation}\\
Our experiments are performed using a thermal gas of potassium-39 atoms, loaded directly from a magneto-optical trap (MOT) into a crossed optical dipole trap. The resulting cigar shaped atom cloud has a temperature of 40 $\,\mathrm{\mu K}$ and $e^{-1/2}$ radii of $10 \,\mathrm{\mu m}$ by $100\,\mathrm{\mu m}$. This should be compared to the characteristic distance between facilitated Rydberg excitations $r_\mathrm{fac}=(C_6/\Delta)^{1/6}$, which for a detuning of $\Delta/2\pi=30\,\mathrm{MHz}$ is $\approx 1.7\,\mathrm{\mu m}$. The peak number of atoms in the $|g\rangle$ state is $1.3 \times 10^{5}$ and the density determined by in-situ imaging is $2.4\times 10^{11}$ cm$^{-3}$. To vary the density while holding all other parameters fixed we reduce the MOT loading time. The lifetime of the atoms in the trap without Rydberg excitation is $\sim4 \,\mathrm{s}$, i.e. much longer than the relevant timescales for the SOC dynamics.\\

\noindent\textbf{B. Excitation laser}\\
To excite the atoms to the $39p_{3/2}$ Rydberg state we use a single photon optical transition at a wavelength of 285 nm. This light is produced by a frequency doubled dye laser delivering up to 200\,mW of single mode light and is frequency stabilised to a high-finesse cavity resulting in an independently measured linewidth of 400 kHz. The excitation beam is aligned parallel to the long axis of the trap and weakly focused to a waist much larger than the size of the atom cloud such that it is practically uniform. We experimentally determine the Rabi frequency $\Omega$ for every individual repetition of the experiment by logging the respective single-shot laser power on a photodiode and employing an independent Rabi frequency calibration based on measuring light shifts induced by the laser via Ramsey interferometry.  \\

\noindent\textbf{C. Numerical simulation of the Langevin equation}\\
\noindent While the Langevin equation~\eqref{eq:langevin} is straightforward to solve in the mean field approximation, in Fig.~\ref{fig:1_temporal} we show exemplary numerical simulations that capture the effects of diffusion and multiplicative noise terms in a one-dimensional setting. For these simulations we make use of the XMDS2 (stochastic) differential equation solver package~\cite{Dennis2013}, assuming a transverse grid size of 128 points and timestep of $2.5\times 10^{-3}$. The noise term is implemented as a zero mean Wiener process with a standard deviation $\propto\sqrt{\rho}$. However, to ensure numerical stability we found it necessary to impose a noise cutoff by setting $\xi=0$ when $\rho<0.0025\,n_0$. For $b=0$ the solutions exhibit an absorbing state phase transition at $n_0=0.39$ and powerlaw scaling consistent with directed percolation universality (in one-dimension $\beta_{DP}=0.276$). For $b\neq 0$ we find that individual timetraces obtained from the full numerical solution are qualitatively very similar to the corresponding mean field solutions. By fitting the numerical results in the same manner as performed for the experimental data we obtain slightly larger effective parameters $\kappa$ and $\Gamma$.\\

\noindent\textbf{D. Comparison of the power-law hypothesis to alternative distributions}\\
\noindent To test whether the avalanche data is indeed described by a power-law distribution we employ the widely used Kolmogorov-Smirnov (KS) test against several alternative distributions, including other heavy tailed distributions (following the definitions in Ref.~\cite{Klaus2011}). The KS statistic is defined as the maximum distance between the cumulative distribution of the empirical data and that of the hypothesized distribution, with small values $\ll 1$ indicating good agreement. In all cases we minimize the KS statistic as a function of the parameters of the hypothesized distribution, restricting the data and the hypothesized distributions to the range $20\leq s\leq 400$. For the data depicted in Fig.~\ref{fig:powerlaw} the obtained KS-test statistics are: 0.015 (power-law), 0.102 (exponential), 0.031 (log-normal), and 0.04 (gamma distribution). This shows that the power-law distribution provides a better fit to the data than the alternative distributions. The power-law exponent $\alpha=-1.38$ found via KS minimization is in excellent agreement with the value obtained via maximum-likelihood estimation.

\normalem
\bibliography{soc}

\appendix

\setcounter{figure}{0}
\makeatletter 
\renewcommand{\thefigure}{S\@arabic\c@figure}

\newpage
{\center\section{Supplementary Information}}
\noindent\subsection{Derivation of the Langevin Equation}
The microscopic dynamics of the driven-dissipative Rydberg ensemble is described by the master equation Eq.~(1M). For realistic system sizes required for SOC, it becomes, however, intractable due to the fast growth of the Hilbert space. In order to reduce this theoretical complexity, we eliminate irrelevant degrees of freedom and map the dynamics to the Langevin equation Eq.~(2M).

{\it Adiabatic elimination of atomic coherences ---} In the presence of strong dephasing $\gamma_{\rm{de}}\gg \Omega$ the evolution of the atomic coherences $\hat{\sigma}_{l}^{gr}, \hat{\sigma}_{l}^{0r}$ is dominated by a rapid dissipative decay towards their time averaged expectation values $\langle \hat{\sigma}_{l}^{\alpha,r}\rangle_T = \frac{1}{T}\int_0^T\text{Tr}\left[\hat{\sigma}_{l}^{\alpha,r}\hat{\rho}\right]dt$, where $T\approx \Omega^{-1}$ is the typical time scale for a facilitated Rabi oscillation, $\text{Tr}$ is the trace over the many-body Hilbert space and $\alpha=g,0$. The coherences are static on many-body time scales and will be adiabatically eliminated by solving
\eq{Eq1}{0\overset{!}{=}\partial_t \langle \hat{\sigma}_{l}^{\alpha,r}\rangle=\text{Tr}\left[\hat{\sigma}_{l}^{\alpha,r}\partial_t\hat{\rho}\right].} Here $\partial_t\hat{\rho}$ is set by Eq.~(1M). After eliminating the coherences, the system evolution is governed by the remaining degrees of freedom, i.e. the average densities 
\eq{Eq2}{
m_l\equiv \text{Tr}\left[
\hat{\sigma}_{l}^{rr}\hat{\rho}
\right], \ n_l\equiv\text{Tr}\left[
(\hat{\sigma}_{l}^{rr}+\hat{\sigma}_{l}^{gg})\hat{\rho}
\right].
}
Their equation of motion is
\eq{Eq2a}{
\partial_tm_l= \text{Tr}\left[
\hat{\sigma}_{l}^{rr}\partial_t\hat{\rho}
\right], \ \partial_tn_l=\text{Tr}\left[
(\hat{\sigma}_{l}^{rr}+\hat{\sigma}_{l}^{gg})\partial_t\hat{\rho}
\right]}
and $\partial_t\hat{\rho}$ is set by Eq.~(1M) and constrained to configurations that fulfill Eq.~\eqref{Eq1}. Explicit evaluation yields
\eq{Eq3}{\partial_t n_l&=&-\Gamma m_l+\xi^n_l,\\
\partial_tm_l&=&\text{Tr}\left(\frac{\Omega ^2 (\Gamma+\gamma_{\rm{de}})(\hat{\sigma}_{l}^{gg}-\hat{\sigma}_{l}^{rr})}{(\Gamma+\gamma_{\rm{de}})^2+4( \hat{V}_l-\Delta)^2}\hat{\rho}\right)-\Gamma  m_l+\xi_l^m.\ \ \ \ \ \ \label{Eq4}
}
The Markovian noise fields $\xi_l^{m,n}$ enforce the non-equilibrium fluctuation relation, which is imprinted by the dissipative environment~\eqref{Eq2}. The statistics of $n_l, m_l$, imprinted by drive and dissipation, are expressed by the vanishing mean $\langle \xi^{n,m}_l\rangle=0$ and non-vanishing variance $\text{var}(\xi_l^{n,m})\neq0$, and the Markovianity, i.e. locality in time and space, of the noise. Their variance is determined by the generalized Einstein relation
\eq{Eq8}{
\text{var}(\xi_l^{m})&=&\partial_t\langle (\hat{\sigma}_{l}^{rr})^2\rangle-2\langle\hat{\sigma}_{l}^{rr}  \partial_t \hat{\sigma}_{l}^{rr}\rangle, \nonumber\\
&=& \tau n_l+(\Gamma+2\tau) m_l+O(m_l^2),
}
and similar for $\xi_l^n$ with $\hat{\sigma}_{l}^{rr}\rightarrow\hat{\sigma}_{l}^{rr}+\hat{\sigma}_{l}^{gg}$. In the limit $\tau\rightarrow 0$, the variance of $\xi_l^m$ is multiplicative in $m_l$ which is a necessary condition for a robust absorbing phase.

Equation~\eqref{Eq4} is a coupled set of differential equations describing the Rydberg population and total remaining population for each atom, satisfying the completeness relation $\hat{\sigma}_{l}^{rr}+\hat{\sigma}_{l}^{gg}+\hat{\sigma}_{l}^{00}=\mathds{1}$. 
Interatomic interactions enter as an effective detuning $\hat{V}_l=\rm{C}_6\sum_{l'\neq l}\hat{\sigma}^{rr}_{l'}/|\mathbf{r}_{l,l'}|^6$, where $|\mathbf{r}_{l,l'}|=|\mathbf{r}_{l}-\mathbf{r}_{l'}|$ is the distance between atom $l$ and $l'$~\cite{Marcuzzi2014}.

In order to expand the trace in Eq.~\eqref{Eq4} in the projection operators $\hat{\sigma}_{l'}^{rr}$, one exploits the fact that $\hat{\sigma}_{l'}^{rr}=\left(\hat{\sigma}_{l'}^{rr}\right)^2$. For an arbitrary function $f$ of the projectors $\hat{\sigma}_{l'}^{rr}$, up to linear order in the projection operators one finds $f(\{\hat{\sigma}_{l'}^{rr}\})=f(0)+\sum_m[f(\hat{\sigma}_{l'\neq m}^{rr}=0,\hat{\sigma}_{m}^{rr}=1)-f(0)]\hat{\sigma}_{m}^{rr}$. Consequently, the operator acting on atom $l$
\eq{Eq5}{
&&\frac{\Omega ^2 (\Gamma+\gamma_{\rm{de}})}{(\Gamma+\gamma_{\rm{de}})^2+4( \hat{V}_l-\Delta)^2}=\underbrace{\frac{\Omega ^2 (\Gamma+\gamma_{\rm{de}})}{(\Gamma+\gamma_{\rm{de}})^2+4\Delta^2}}_{
=\tau}+ \nonumber\\ 
&&\sum_{l'}\left[\frac{\Omega ^2 (\Gamma+\gamma_{\rm{de}})}{(\Gamma+\gamma_{\rm{de}})^2+4(\Delta-\rm{C}_6|\mathbf{r}_{l,l'}|^{-6})^2}-\tau\right]\hat{\sigma}_{l'}^{rr}+... ,\,\,\,\,\label{Eq5}
}
followed by higher order products of projectors (i.e. terms $\sim \hat{\sigma}_{l'}^{rr}\hat{\sigma}_{m}^{rr}$). The first term on the right hand side describes single particle excitations with rate $\tau$, while the second term describe the facilitated (de-)excitation of atom $l$ by another atom $l'$ in the Rydberg state.  This describes a Lorentzian peaked at the facilitation radius $|\mathbf{r}_{l,l'}|=(\rm{C}_6/\Delta)^{1/6}\equiv r_{\text{fac}}$ and deviates considerably from zero only for $|\mathbf{r}_{l,l'}|\in [r_{\text{fac}}-\Delta r_{\text{fac}}, r_{\text{fac}}+\Delta r_{\text{fac}}]$ with $\Delta r_{\text{fac}}=r_{\text{fac}}\frac{\Gamma+\gamma_{\rm{de}}}{12\Delta}$. Introducing a projector $\Pi_{ll'}$ with $\Pi_{ll'}=1$ if $|\mathbf{r}_{l,l'}|\in [r_{\text{fac}}-\Delta r_{\text{fac}}, r_{\text{fac}}+\Delta r_{\text{fac}}]$ and zero elsewhere, Eqs.~\eqref{Eq4}-\eqref{Eq5} yield 
 \eq{Eq6}{
\partial_tm_l&=&\Big(\tau+\sum_{l'}\frac{\Omega^2\Pi_{ll'}m_{l'}}{\Gamma+\gamma_{\text{de}}}\Big)(n_l-2m_l)-\Gamma  m_l+\xi^m_l.\ \ \ }

Equation \eqref{Eq6} provides a good approximation to the facilitation rate assuming the excitation density is small, but overestimates the true facilitation rate when there is several Rydberg excitations in proximity to state $l$ (due to truncating the expansion \eqref{Eq5} at first order). An exact computation of the facilitation radius for $w\ge1$ excited states inside a single shell shows that it grows as $r_{R}^{(w)}=w^{1/6}r_R$ (in $d=3$ dimensions). For a homogenous distribution of atoms, this yields a facilitation rate that grows proportional to $\sqrt{w}$, which is not a severe correction compared to the $\propto w$ growth predicted by Eq.~\eqref{Eq6} if one bears in mind the largely suppressed off-resonant excitation rate. 
More than a single excitation inside the facilitation radius, i.e., $w>1$, can only be realized via additional spontaneous excitation events. The probability for $w>1$ is a factor of $\mathrm{O}(10^{-4})$ smaller compared to $w\leq 1$ and will have no impact on the dynamics.

{\it Continuum Limit ---} For the reported experiments the atoms are free to move on the timescale of the slow SOC dynamics. However, the diffusion time scale (set by the temperature) for distances of the order of the facilitation radius $O(r_{fac})$ is about one order of magnitude slower than the inverse Rabi frequency. This justifies an effectively static model for the external degrees of freedom while ensuring that the the ground state density remains approximately homogenous on length scales compared to the facilitation radius. Thus we can coarse grain the dynamics by averaging the densities over facilitation shells
\eq{Eq9}{
\rho({\bf r},t)\equiv\mathcal{N}\hspace{-2mm}\sum_{|{\bf r}_l-{\bf r}|\le r_{\text{fac}}}\hspace{-4mm}m_l,\ \ \   n({\bf r},t)\equiv\mathcal{N}\hspace{-2mm}\sum_{|{\bf r}_l-{\bf r}|\le r_{\text{fac}}}\hspace{-4mm}n_l,\ \ \
}
where $\mathcal{N}=(\tfrac{4\pi}{3} r_{\text{fac}}^3)^{-1}$ is the normalization volume.
 
This coarse-graining procedure modifies the completeness relation compared to the single atom case. An excited atom facilitates excitations at the border of the facilitation shell but blocks the excitation for any atoms within the shell. Decay of a Rydberg excitation to a removed state thus removes the blockade constrant on the remaining ground state atoms. At the scale of the facilitation radius, the averaging procedure \eqref{Eq9} yields the effective rate of decay into removed states is $\Gamma \rightarrow b\Gamma$ and adds an effective decay rate back to the ground state $\Gamma(1-b)$, where $b=\rho({\bf r},t)/n({\bf r},t)\approx \mathrm{const.}$ Defining $n_t\equiv n({\bf r},t), \rho_t\equiv\rho({\bf r},t)$, the averaged densities evolve as
\eq{Eq10}{
\partial_t n_t&=&-b\Gamma \rho_t+\xi_t,\\
\partial_t \rho_t&=&-\Gamma \rho_t+\xi_t+(n_t-2\rho_t)\Big(\tau+\tfrac{\Omega^2}{\Gamma+\gamma_{\text{de}}}\mathcal{M}(\rho_t)\Big).\label{Eq11}
}
The averaged noise $\xi_t$ remains Markovian in time and space with variance $\text{var}(\xi_t)=\tau n_t+(\Gamma+2\tau)\rho_t$. The nonlinearity $\mathcal{M}(\rho_t)$ is obtained from the execution of the density averaging \eqref{Eq9} in the sum $\sim \sum_{l'}\Pi_{ll'}m_{l'}$ in Eq.~\eqref{Eq6}. It is a non-local function in space and has to be read as $\rho_t\mathcal{M}(\rho_t)=\rho({\bf r}, t)\int_{{\bf r-r}'\in S_{\text{fac}}}\rho({\bf r}', t)$ with $S_{\text{fac}}=[r_{\text{fac}}-\Delta r_{\text{fac}}, r_{\text{fac}}+\Delta r_{\text{fac}}]$ being the facilitation shell. Taking advantage of the smooth densities for $|{\bf r}_{ll'}| =r_{\text{fac}}$ we can perform a Taylor expansion of $\rho({\bf r}', t)$, yielding
\eq{Eq15}{
\mathcal{M}(\rho_t)=\mathcal{M}(1)\rho_t+\frac{\mathcal{M}({\bf r}^2)}{2}\nabla^2\rho_t+O(\nabla^4\rho_t),
}
where odd derivative terms vanished due to isotropy in space. The factors $\mathcal{M}(1)=\int_{{\bf r}\in S_{\text{fac}}} 1 $ and $\mathcal{M}({\bf r}^2)=\int_{{\bf r}\in S_{\text{fac}}} {\bf r}^2 $ are the averages of $1,{\bf r}^2$ along the facilitation shell.

Including thermal diffusion with diffusion constant $D_T$ caused by the thermal motion of the atoms this yields the final form of the Langevin equation
\eq{Eq17}{
\partial_t n_t&=&D_T\nabla^2n_t-b\Gamma \rho_t+\xi_t,\\
\partial_t \rho_t&=&(D\nabla^2+\kappa n_t-\Gamma-2\tau)\rho_t-2\kappa\rho_t^2+\tau n_t+\xi_t.\label{Eq18}\ \ \ \ 
}
In the experiment $\Gamma/\tau\approx10^4$ and $n_t/\rho_t\approx 20$, justifying $\Gamma+2\tau\rightarrow\Gamma$ and $n_t+\rho_t\rightarrow n_t$. Together with the numerical simulations this yields the estimate $b\approx0.05$. Furthermore, the theoretical derivation predicts $\kappa=\frac{\Omega^2}{\Gamma+\gamma_{\text{de}}}\mathcal{M}(1)\approx\frac{2\pi\Omega^2}{3\Delta}r_{\text{fac}}^3$ and ${D\approx D_T+n_t\frac{\pi\Omega^2}{3\Delta}r_{\text{fac}}^5}$. Assuming van der Waals interactions $r_\mathrm{fac} = (C_6/\Delta)^{1/6}\approx 1.7\,\mathrm{\mu m}$. The diffusion of excitations $\sim D$ is primarily governed by fast facilitation of neighboring atoms and only marginally affected by temperature, i.e. $D_T\ll n_t\frac{\pi\Omega^2}{3\Delta}r_{\text{fac}}^5$. The numerical simulations of Eq.~\eqref{eq:langevin} presented in Fig.~\ref{fig:1_temporal}d for $D_T=0$ show that for experimentally relevant parameters $n_t$ remains mostly homogeneous during the evolution since the diffusion of $\rho_t$ is sufficiently fast compared to the effective loss rate $b\Gamma$. Thus for the conditions of the experiment atomic motion has very little impact on the qualitative SOC behaviour. However this might change in some lattice systems for example, where additional geometric constraints could have a more dramatic effect on the SOC dynamics.

The structure of the Langevin equations \eqref{Eq17}, \eqref{Eq18} is obtained from the discussed, controlled coarse graining procedure and is insensitive to minor variations of the microscopic details. The effective parameters $D, \Gamma, \kappa, b, \tau$, however, can be influenced by such variations, that may include disorder, atomic motion and cooperative excitation processes, in the present setup. For this reason, the predicted values above only serve as a rough guide and in order to compare theoretical predictions with the experimental results we fit the data to mean field solutions of Eq.~\eqref{Eq17} to consistently determine the relevant parameters of the model.

\noindent\subsection{Detuning dependence and further evidence for non-equilibrium universality}

\begin{figure}[!t]
	\centering 
	\hskip -5pt\includegraphics[width = 0.75\columnwidth]{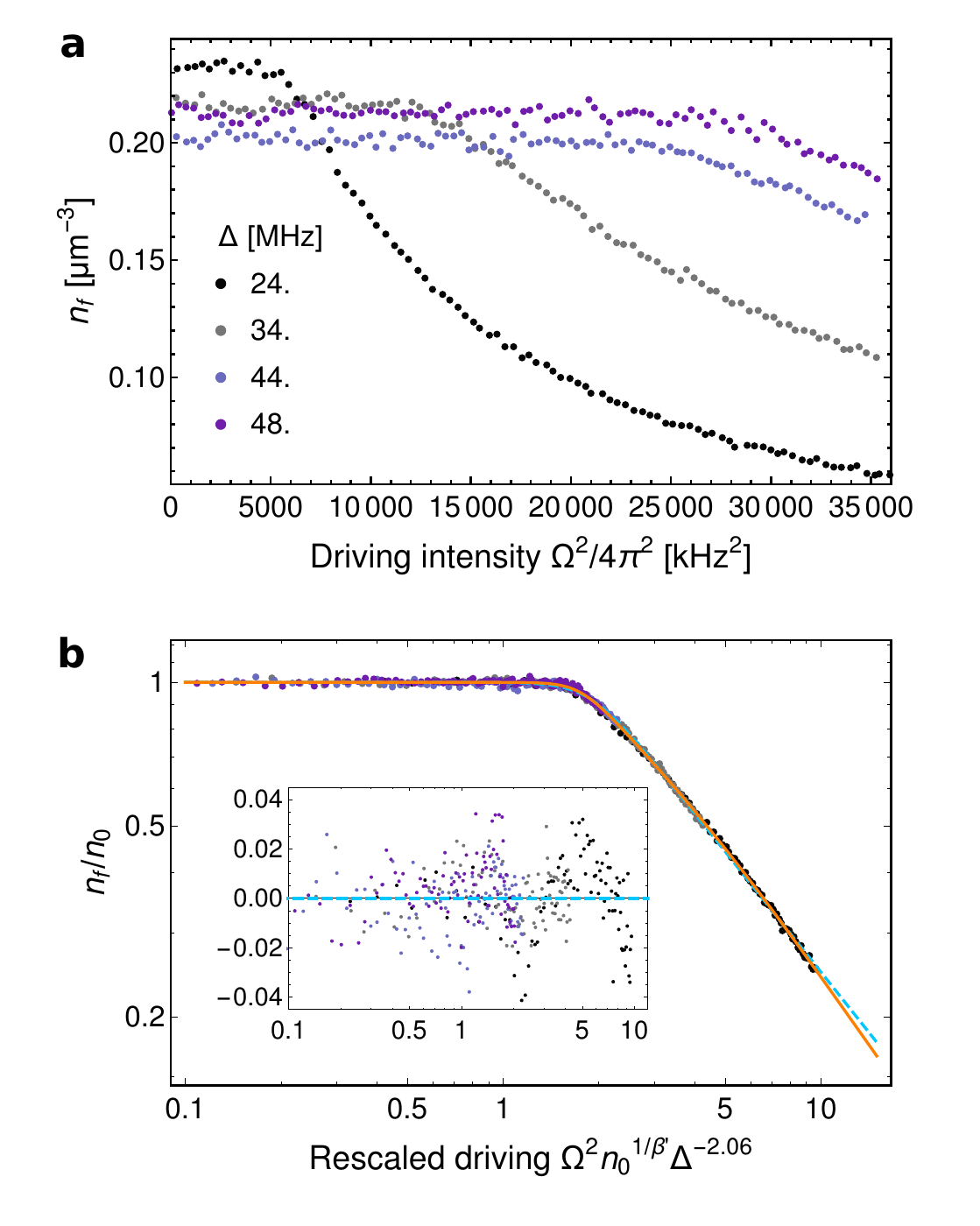}
	\caption{\textbf{Further evidence for non-equilibrium universality} (a) Stationary state density $n_f$ measured at $t = 10\,\mathrm{ms}$ as a function of $\Omega^2$ and for different detunings $\Delta$. (b) The same data with rescaled axes to achieve full data collapse, revealing the scaling function (with fit shown by the dashed blue line) for the stationary density $n_f$. The inset shows the normalised residuals between the rescaled data and the fitted scaling function. The dashed blue line corresponds to the simple scaling function used in the main text, while the solid orange line is a generalised scaling function which reproduces the asymptotic scaling form more accurately.} 
	\label{fig:S1_datacollapse}
\end{figure}

In the following we present additional evidence for the universal nature of the self-organised stationary state. For this we performed additional measurements of the stationary density as a function of the driving intensity but for different detunings of the excitation laser as shown in Fig.~\ref{fig:S1_datacollapse}. Each dataset shows qualitatively similar behavior to that presented in Fig.~\ref{fig:2_perceo}, clearly showing the transition from an absorbing phase to a self-organising active phase. However this data also clearly shows that the location of the critical point depends on the laser detuning. 
  
To further analyze this data we apply the scaling ansatz $n_f = n_0 F(\Omega^2 \Delta^d n_0^{1/\beta'})$, where $\beta'=0.869$ and we have included as a new parameter the detuning rescaling exponent $d$. For $d=-2.06(1)$ the data collapses once again onto a single universal curve. In this way we determine the $\kappa\propto \Omega^2/\Delta^{2.06}$ dependence of the spreading parameter, used elsewhere in the paper to compare the data with mean field theory.

Before analysing the scaling properties of the rescaled data, very careful inspection shows that it has a slightly different form to the scaling function $F(x)$ used to describe the data in Fig.~\ref{fig:2_perceo}b. This is evidenced by the fit to $F(x)$ shown as a blue dashed line in Fig.~\ref{fig:S1_datacollapse}b. The deviation is most apparent in the normalized fit residuals (Fig.~\ref{fig:S1_datacollapse}-inset) which, in contrast to Fig.~\ref{fig:2_perceo}b, exhibits some structure (e.g. the inverted U-shape of the black datapoints). Unless properly accounted for, this deviation between the scaling form of the data and the heuristic scaling function causes a systematic error in the determination of the critical scaling exponent. To rectify this, we model the detuning dependent data by a generalized scaling function $F'(x)=[1+(x/x_a)^{v \alpha}+(x/x_c)^{v \beta}]^{-1/v}$, where the newly introduced parameters $x_a<x_c$ and $\alpha<\beta$ empirically describe power-law scaling for intermediate driving intensities. In the asymptotic regime $x\gg x_c$, the scaling function once again reduces to a power law $n_f/n_0\propto x^{-\beta}$.

\noindent\subsection{Critical response}
\begin{figure}[!t]
	\centering 
	\includegraphics[width = 0.56\columnwidth]{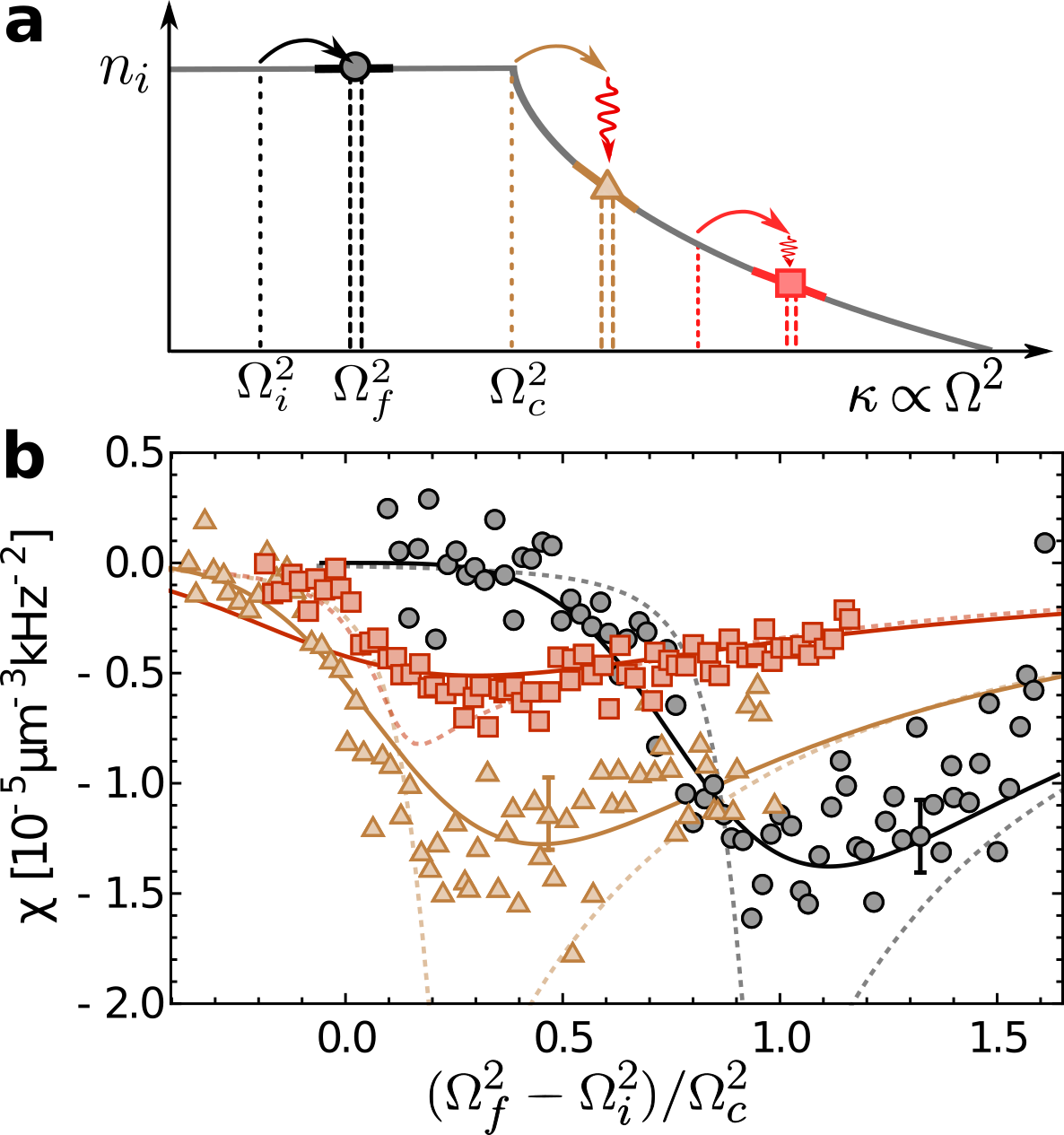}
	\caption{\textbf{Response of the self-organised critical state to external perturbations.} (a) Sketch of the experimental procedure used to measure the susceptibility $\chi = \mathrm{d}n_f/\mathrm{d}(\Omega_f^2)$ by quenching the spreading parameter $\kappa\propto \Omega^2$ across the absorbing state phase transition. (b) Experimental data corresponding to three different initial conditions corresponding to the absorbing phase (black circles), critical (brown triangles) and active phase (red squares). The solid lines correspond to predictions based on the experimentally determined scaling function and dotted lines correspond to mean field predictions. For reference we show two representative error bars corresponding to the standard error of the mean.} 
	\label{fig:3_sus}
\end{figure}

As additional evidence for the system reaching a critical state, we have investigated the gapless response of the stationary state following a parameter quench. Assuming the SOC state is indeed an attractor for the dynamics we expect that small perturbations, e.g. a sudden change of the spreading parameter $\kappa$, should trigger avalanche like processes that eventually bring the system back to a new critical state corresponding to a lower stationary density. On the other hand, if the system evolves to a state which is deep within the absorbing phase, then avalanches can only be triggered by perturbations larger than a threshold value corresponding to a non-zero dynamical gap. To measure this response we start from the stationary state (reached after $t=10\,$ms) corresponding to different driving intensities $\Omega_i^2$ (sketched in Fig.~\ref{fig:3_sus}a). We then perturb the system by quenching the driving intensity to a new value $\Omega_{f1}^2$ and then wait for a further 10 ms before measuring the new stationary density. The whole procedure is then repeated with a slightly larger final driving intensity $\Omega_{f2}^2\approx \Omega_{f1}^2+(2\pi\times 50~\mathrm{kHz})^2$. From these two measurements we estimate the susceptibility $\chi=dn_f/d(\Omega_f^2)=[n_f(\Omega_{f1}^2)-n_f(\Omega_{f2}^2)]/(\Omega_{f1}^2-\Omega_{f2}^2)$ where $\Omega_f=(\Omega_{f1}+\Omega_{f2})/2$. 

Figure~\ref{fig:3_sus} shows the measured susceptibility as a function of $\delta=(\Omega_{f}^2-\Omega_{i}^2)/\Omega_c^2$ for three different initial conditions corresponding to $\Omega_i<\Omega_c$ (absorbing), $\Omega_i\approx \Omega_c$ (critical) and $\Omega_i>\Omega_c$ (active). For each of these initial conditions we observe pronounced minima in $\chi$ corresponding to the strongest system response. We interpret the leading edge on the left side of each minimum as the point where the perturbation is sufficient to bring the system back to the active phase, thereby triggering avalanche-like dynamics and extra loss. When starting deep in the absorbing phase (black circles) the onset occurs at a large value of $\delta$, which is a measure of the non-zero dynamical gap. In contrast, the onsets for critical (brown triangles) and active (red squares) initial states both coincide at $\delta\approx 0$ within the experimental resolution. We can compare this data to a prediction of the susceptibility obtained from a derivative of the experimentally determined scaling function using $\beta=0.910$. The scaling function predictions (solid lines in Figure~\ref{fig:3_sus}) are in good agreement with the data, whereas mean field predictions (dotted lines) systematically fail to capture the widths and heights of the peaks. We note however that starting from the initial active state, the measured response is narrower and slightly stronger than the scaling function prediction. This is further evidence that the system evolves towards a state, which is sharply concentrated at the critical point rather than a statistical mixture of many different accessible macro states. From these experiments we confirm that when starting from a supercritical state (irrespective of the precise value of $\Omega_i> \Omega_c$), the system self-organises to a critical state which is characterised by a vanishing excitation gap [underpinning signature (iii)].

\noindent\subsection{Role of trap inhomogeneities and residual coherence}

Fitting the generalized scaling function to the rescaled data yields $\beta=0.95(3)$ where the larger statistical uncertainty reflects the fact that the generalized model function has more parameters. This is close to the value for $\beta=0.910(4)$ determined from the density dependent data in the main text. Refitting the density dependent data with the generalized scaling function yields $\beta=0.920(7)$. This shows that, while the full form of the scaling function is not universal, data taken under very different conditions concerning initial densities and detuning of the driving field do in fact share a common universal critical exponent describing the asymptotic scaling regime.

We can also rule out possible modifications to the scaling behavior due to other experimental details such as the inhomogeneous density or residual effects of quantum coherence: (i) Inhomogeneities - in the experiment the atoms are laser-trapped in a cylindrical geometry of finite diameter and length, causing a nearly homogeneous density distribution in the trap center and smooth variation of $n_t$ at the boundaries. $n_t$ follows smoothly the Gaussian trapping profile of the lasers. To estimate the impact of inhomogeneities, on this basis we now study a local density approximation for the Langevin equation. In this approximation, $\rho({\bf r},t)$ experiences a constant background density $n({\bf r}, t)=\tilde{n}({\bf r},t)I({\bf r})$ at each point in space ${\bf r}$, which is modulated by the trapping profile $I({\bf r})$, whereas $\tilde{n}({\bf r},t)$ only incorporates fluctuations due to the coupling to $\rho_t$. An appropriate mean-field theory considers $\rho_t=V^{-1}\int_V \rho({\bf r},t)$ as the spatially averaged density over the system volume $V$ and $\tilde{n}_{t=0}=n_0$ in the absence of fluctuations. The corresponding, spatially averaged SOC line is located at $\Omega^2_c\sim \kappa_c=\frac{\Gamma}{n_0}\int_V \frac{1}{I({\bf r})}\sim n_0^{-1}$, demonstrating that the mean-field exponent $\beta=1$ is not modified by the inhomogeneous geometry. 

(ii) Quantum coherence - the evolution of the density averages \eqref{Eq3}, \eqref{Eq4} is real and linear in time, which maps the final Langevin equation to a stochastic differential equation for classical processes. It incorporates strong, classical correlations between different atoms but lacks the possibility for long range coherence. Coherence between different atoms can be built in systematically by replacing the adiabatic elimination in Eq.~\eqref{Eq1} with the exact solution, which amounts to a shift $\Gamma\rightarrow \Gamma+\partial_t$ in \eqref{Eq4}. To leading order it introduces a coherent contribution $(\Gamma+\gamma_{\text{de}})^{-1}\partial_t^2m_l$ to the RHS of \eqref{Eq4}. Analogously to a damped harmonic oscillator, this evolution is observable on timescales $t(\Gamma+\gamma_{\text{de}})\le 1$ and washed out on larger times scales, i.e. on the relaxation towards the SOC steady state. Fast coherent processes might modify the parameters $\kappa, D, \tau$ but not the structure of the Langevin equation.

\end{document}